\documentclass[sn-apa]{sn-jnl}

\usepackage{graphicx}%
\usepackage{multirow}%
\usepackage{amsmath,amssymb,amsfonts}%
\usepackage{amsthm}%
\usepackage{mathrsfs}%
\usepackage[title]{appendix}%
\usepackage{xcolor}%
\usepackage{textcomp}%
\usepackage{manyfoot}%
\usepackage{booktabs}%
\usepackage{algorithm}%
\usepackage{algorithmicx}%
\usepackage{algpseudocode}%
\usepackage{listings}%
\usepackage{cite}%
\usepackage{setspace}%
\usepackage[none]{hyphenat}%
\usepackage{natbib}
\usepackage{xurl}
\usepackage{cleveref}
\usepackage{lmodern}%
\usepackage{lineno}%

\usepackage[skip=10pt]{parskip}

\begin{document}
\sloppy{}
\title[Article Title]{A Comparison between Financial and Gambling Markets}


\author*{\fnm{Haoyu} \sur{Liu}}\email{hl215@st-andrews.ac.uk}

\author{\fnm{Carl} \sur{Donovan}}\email{crd2@st-andrews.ac.uk}
\author{\fnm{Valentin} \sur{Popov}}\email{vmp@st-andrews.ac.uk}

\affil{\orgdiv{School of Mathematics and Statistics}, \orgname{University of St Andrews}, \orgaddress{\city{St Andrews}, \postcode{KY16 9SS}, \state{Fife}, \country{United Kingdom}}}
\abstract{Financial and gambling markets are ostensibly similar and hence strategies from one could potentially be applied to the other. Financial markets have been extensively studied, resulting in numerous theorems and models, while gambling markets have received comparatively less attention and remain relatively undocumented. This study conducts a comprehensive comparison of both markets, focusing on trading rather than regulation. Five key aspects are examined: platform, product, procedure, participant and strategy. The findings reveal numerous similarities between these two markets. Financial exchanges resemble online betting platforms, such as Betfair, and some financial products, including stocks and options, share speculative traits with sports betting. We examine whether well-established models and strategies from financial markets could be applied to the gambling industry, which lacks comparable frameworks. For example, statistical arbitrage from financial markets has been effectively applied to gambling markets, particularly in peer-to-peer betting exchanges, where bettors exploit odds discrepancies for risk-free profits using quantitative models. Therefore, exploring the strategies and approaches used in both markets could lead to new opportunities for innovation and optimization in trading and betting activities.}



\maketitle

\newpage
\doublespacing
\section{Introduction}\label{sec1}
Recently, a growing number of professional theoretical and empirical research has illuminated surprising parallels between financial and gambling markets \citep{borna1987gambling, arthur2016conceptual, cox2020compulsive, weidner2022gambling}. Both systems, financial markets and gambling markets \citep{schwartz2006roll}, have a long history. Financial markets facilitate the exchange of assets, including stocks and bonds \citep{valdez2016introduction} and gambling markets involve betting on uncertain outcomes \citep{borna1987gambling}. However, the literature on models and strategies for gambling markets is relatively limited compared to the widespread literature available on financial markets. Several factors contribute to this discrepancy, including the inherent complexity of the markets \citep{blau2020gambling} and the regulatory environment surrounding gambling \citep{weidner2022gambling}. Gambling markets represent a simplified form of financial markets \citep{williams1999information}. Therefore, the models and strategies proven effective in financial markets could potentially be adapted and applied to gambling markets \citep{williams2023financial}. A deeper comparison between these two markets could provide valuable insights into the potential transferability of financial market approaches to the gambling arena, and may uncover opportunities for developing new and innovative strategies for gambling. 

Historically, financial markets could be traced back to ancient civilizations including Mesopotamia and Egypt. During this period, merchants and traders engaged in the exchange of goods and early forms of banking \citep{carmona2007accounting}. However, it was during the $17^{\text{th}}$ and $18^{\text{th}}$ centuries that modern financial markets began to take shape. In 1602, the world's first official stock exchange, the Amsterdam Stock Exchange, was established, marking a significant milestone in the development of financial markets \citep{petram2011world}. After World War II, the financial markets experienced profound transformations, largely influenced by the Bretton Woods agreement of 1944. This agreement introduced a new international monetary system with fixed exchange rates pegged to the U.S. dollar \citep{dooley2004revived}. Consequently, financial markets became more globalized, and multinational corporations emerged as prominent players in the international economic landscape. The eventual collapse of the Bretton Woods system paved the way for significant technological advancements in financial markets. Electronic trading and computer-based systems emerged, revolutionizing the trading process by making it faster, more efficient, and accessible to a broader range of participants \citep{gomber2018fintech}. The introduction of computers and the internet played a pivotal role in this transformation. In recent years, financial markets encompass a diverse array of assets, including stocks, bonds, options, currencies, and derivatives. These markets are deeply interconnected on a global scale, with exchanges, banks, investment firms, and individual investors all actively participating in trading and investment activities \citep{valdez2016introduction}. As a result, the modern financial landscape is characterized by dynamic and fast-paced interactions among various stakeholders across the world.

Similar to financial markets, gambling markets also have a long history. The origins of gambling could be traced back to the Paleolithic period \citep{schwartz2006roll}. The first recorded modern casino opened in 1638 in Italy \citep{soligo2021playing}. Since then, gambling has been a popular form of entertainment, with horseracing emerging as one of the most prevalent betting themes \citep{munting1996economic, huggins2014flat}. In the late 1990s and early 2000s, the internet brought significant changes to the gambling markets, which online gambling gained immense popularity \citep{mayra2015mobile}. Online gambling markets have rapidly expanded, progressively being favored by gamblers \citep{philander2014online}. In the past 10 years,  the peer-to-peer online gambling market has been founded and become increasingly popular \citep{rieche2007peer}. Taking Betfair as an example, the platform has revolutionized the gambling experience, allowing individuals to directly interact and place bets against each other \citep{cameron2009you}. Therefore, the evolution of gambling markets mirrors that of financial markets in some respects, with historical roots and significant shifts in recent years due to technological advancements and changing consumer preferences \citep{williams1999information}.

As highlighted, both markets have seen the prevalence of peer-to-peer markets conquering the traditional domain of large companies, such as stocks, options and the provision of betting odds. Tim Berners-Lee’s original vision for the World Wide Web created the concept of P2P computing or networking, characterized by distributed application architecture that distributes tasks among equipotent peers. The architecture underlying P2P systems gained substantial traction following the release of the file sharing system Napster in 1999, which significantly popularized this framework \citep{barkai2001peer}. These participants hold equal privilege within the application’s framework. After that, enabling users to establish a virtual network detached from physical constraints, operating independently of administrative authorities or restrictions \citep{steinmetz2005peer}. This has made P2P systems a significant area in various fields, including financial markets and gambling markets, which are likely to reshape traditional market dynamics and foster greater inclusivity and accessibility for participants. The peer-to-peer system has revolutionized the way transactions and interactions occur in these markets, providing individuals with more direct and unrestricted access to financial instruments and betting opportunities \citep{bebbington2017studies}. This decentralization of trading activities has democratized the process, empowering users to engage in peer-to-peer transactions without relying on intermediaries or centralized authorities.

In this sense, the present study does not merely catalogue descriptive similarities between financial and gambling markets, but seeks to establish a conceptual bridge that enables systematic cross-market knowledge transfer. In particular, peer-to-peer gambling markets exhibit core microstructural characteristics, such as decentralized price formation, continuous order matching, back-lay spreads, and direct participant interaction, which are fundamentally analogous to those observed in modern financial exchanges. These shared institutional and operational features suggest that peer-to-peer betting markets can be interpreted not simply as recreational platforms, but as functionally comparable speculative trading environments. Therefore, the view adds to a more unified understanding of speculative behavior under uncertainty and forms a theoretical foundation for integrating gambling markets into broader financial market analysis. The study therefore locates peer-to-peer gambling markets within a financially interpretable framework for building transferable models, strategies and analytical tools that transcend conventional disciplinary boundaries.

Therefore, the primary objective of this paper is to explore the similarities between financial and gambling markets, considering traditional markets as well as peer-to-peer markets. Despite the growing literature comparing financial and gambling markets, existing studies primarily remain descriptive and fragmented, with limited integration of market microstructure, behavioral mechanisms, and transferable quantitative strategies. Moreover, peer-to-peer gambling markets have rarely been examined within a unified financial market framework. This study addresses this gap by providing a structured comparative analysis across some components and by explicitly evaluating the transferability of financial trading concepts to modern betting exchanges. The key components of financial markets primarily include trading (for instance, participants, instruments, and intermediaries) and regulatory institutions \citep{kidwell2016financial}. This paper focuses on trading since the regulatory institutions have previously been compared \citep{weidner2022gambling}. The comparison will be conducted across five key aspects, which are the platform, product, procedure, participant, and strategy. 
By using the 4W1H (i.e., what, when, where, who, and how) methodology \citep{lee20214w1h}, the authors argue that these five components comprise the essence of trading.

By examining these aspects, the paper aims to gain insights into the commonalities shared by these markets and identify any potential transferability of models and strategies between them. The study illuminates the interconnectedness and similarities between financial and gambling markets, encompassing peer-to-peer markets as well. The contribution of this research study is to formalize the peer-to-peer gambling markets into quasi-financial trading environments and a systematic framework for the cross-market strategy transfer. This offers potentially valuable insights for investors, traders, and participants from both markets, but especially for financial analysts with an interest in understanding gambling markets. In addition, if strategies and approaches are examined in these markets, more innovative and sophisticated trading and betting activities could be possible. A clear process of screening and categorization was adopted for the selection of the literature. The initial literature pool was generated through targeted keyword searches in Google Scholar (for gambling, finance, and peer-to-peer) and searches in the leading gambling and finance journals. After removing duplicates and non-peer-reviewed sources, studies were then ranked according to relevance, methodological quality and the relevance to the overall concept for cross-market comparison. The chosen publications were analyzed systemically for five analytical dimensions, including platform, product, procedure, participant and strategy, which represent the structural skeleton of the review. This review uses a conceptual synthesis rather than chronological survey, aiming to derive structural insights for the applicability of trading strategies to other markets. A total of 263 publications were screened at an initial stage. We included studies if they examined the effect of the market structure, trading behavior or transferability of the strategy between gambling and financial settings. The final sample included 125 academic paper which published between 1968 to 2026. Each section studies one key aspect and describes both markets and their similarities. Section 2 introduces the similarities between platforms in both markets, including trading and speculation. Section 3 describes the comparable characteristics of products, including risk. Section 4 discusses procedural similarities, while Section 5 examines the participants' purposes. Section 6 explores strategies from financial markets, including statistical arbitrage, which could be applied to gambling markets. Finally, Section 7 presents the conclusions and potential applications.

\section{The Platform}\label{sec2}

\subsection{Traditional Financial Platforms}
In financial markets, there are various types of platforms, including stock exchanges and future exchanges. Both these platforms fall under the category of traditional exchanges in the financial market and involve various agents, including stock broker agents \citep{valdez2016introduction}.

Firstly, a stock exchange acts as a centralized marketplace where buyers and sellers convene to trade financial securities including stocks, bonds, and derivatives. Stock exchanges offer companies an opportunity to raise capital by issuing shares and allow investors to buy and sell those shares \citep{valdez2016introduction}. These transactions are matched electronically based on price and other relevant criteria \citep{becker1993automated}. To ensure fair and orderly trading, stock exchanges implement rules and regulations \citep{gadinis2006markets}. They closely monitor trading activities, investigate irregularities, and may suspend or delist companies that fail to adhere to these rules. This ensures that investors can quickly and fairly buy or sell securities. Stock exchanges often calculate and maintain various market indices, such as the Dow Jones Industrial Average (DJIA) or the S\&P 500. Some well-known stock exchanges globally include the New York Stock Exchange (NYSE) and the Nasdaq in the United States, the London Stock Exchange (LSE) in the UK, and the Shanghai Stock Exchange (SSE) in China. 

Secondly, futures exchange is another significant exchange in the financial markets. It refers to a centralized marketplace where futures contracts are traded, also known as a ``commodity futures exchange'' \citep{valdez2016introduction}.

\subsection{Traditional Gambling Platforms}
Traditional gambling markets encompass the conventional form of gambling, where individuals place bets on the outcomes of various events or sports competitions \citep{gainsbury2015betting}. While in the traditional financial markets there are agents, traditional gambling markets also involve agents, known as registered representatives \citep{sauer1998economics}. These markets have a long-standing history and are commonly associated with physical bookmakers or betting shops. Within traditional gambling markets, bookmakers play a pivotal role as intermediaries, setting the odds and accepting bets from individuals \citep{cain2003favourite}, for example well-known bookmakers in the UK include Ladbrokes and Betfred. One distinguishing aspect of traditional gambling markets is their reliance on physical locations, such as betting shops or bookmaker establishments, where bettors can personally visit to place their bets. These locations often foster a social environment for individuals interested in gambling and sports \citep{flood2000gambling}. 

Popular sports events, including football and horse racing, constitute the primary focus of traditional betting markets \citep{humphreys2012bets}. Participants engage by placing bets on specific outcomes, such as predicting the winner of a football match. Traditional gambling markets are distinguished by their reliance on fixed odds, where the bookmaker establishes the odds for each bet at the time of placement. These odds remain stable, irrespective of any subsequent changes in the market \citep{labrie2007assessing}. This provides participants with clarity and certainty about the potential payout they might receive if their bet turns out to be successful.

\subsection{Online Trading Platforms}
In recent years, the swift progress of technology has triggered a notable increase in the prevalence of online financial trading platforms. These platforms have revolutionized the way individuals participate in trading, eliminating the need for traditional brokers or agents \citep{chen2016advertising}. Traders can directly access real-time market data, execute trades, and manage their positions through these electronic platforms \citep{pavlou2007understanding}. For example, E-trade offers an online trading platform enabling individuals to trade futures contracts without the involvement of a broker \citep{alam2019impact}. Similarly, in China, Shanghai International Energy Exchange (INE) provides an electronic platform for trading crude oil futures contracts \citep{lv2020crude}. 

Besides financial trading, online platforms have also revolutionized the world of online gambling. Peer-to-peer gambling platforms, such as online sportsbooks or betting exchanges, have emerged. These platforms offer similar functionalities to traditional betting markets but provide the convenience of accessibility from a computer or mobile device \citep{gainsbury2017blockchain}. Taking Betfair as an example, the UK company operates the world's largest betting exchange \citep{gonccalves2019deep}. It enables individuals to place bets on various events, including sports and politics. What sets Betfair apart from traditional bookmakers is that it enables users to bet against each other instead of betting against the bookmakers. Participants can either back an outcome they believe will happen or take the opposite position and lay bets against it. Betfair acts as an intermediary, facilitating the matching of bets between users and charging a commission on the winnings \citep{casadesus2019platform}. 

In both markets, these online peer-to-peer platforms have significantly transformed their respective industries, providing individuals with the freedom to engage in trading and gambling without the need for physical presence or intermediaries. With the convenience of access and real-time data availability, the popularity of these platforms is expected to continue growing in the coming years.

\subsection{Comparison between Financial and Gambling Platforms}
While both financial and gambling platforms involve trading and speculation, there are fundamental distinctions between them. Financial platforms, such as stock exchanges, primarily focus on securities and investments within the financial markets \citep{gadinis2006markets}. In contrast, platforms including Betfair operate in the domain of online gambling and betting \citep{cameron2009you}. Moreover, financial platforms including stock exchanges are subject to stringent regulations and are designed to facilitate long-term investments and capital raising \citep{valdez2016introduction}, whereas Betfair betting exchange is oriented towards short-term betting opportunities \citep{gonccalves2019deep}.

Despite the explored disparities, similarities could be found between financial platforms and gambling platforms, particularly when it comes to peer-to-peer markets. Firstly, both online futures exchanges and Betfair betting exchange operate through online platforms, allowing participants to conveniently and remotely access them from their computers or mobile devices \citep{krotov2017internet}. Secondly, both platforms utilize an exchange model, where participants can either place bets or trade contracts directly with each other \citep{franck2010prediction}. Thirdly, market liquidity is a critical factor for both platforms, as higher liquidity ensures sufficient counterparties for participants to enter and exit positions at desired prices \citep{awrey2014limits}. Finally, both platforms provide risk management tools. In futures exchanges, participants can use techniques including hedging and stop-loss orders to manage their risk exposure \citep{papaioannou2006exchange}. Similarly, Betfair betting exchange offers features including cash-out and in-play betting, allowing participants to lock in profits or minimize losses during an event \citep{gainsbury2012internet}. 

Additionally, both types of platforms involve speculation over future events \citep{irwin2011index, lopez2018controlling}. While participants in financial platforms engage in speculative activities by predicting price movements and market trends \citep{irwin2011index}, those in gambling platforms speculate on the outcomes of various events \citep{lopez2018controlling}. Therefore, while financial and gambling platforms have distinct focuses and regulations, they share similarities in terms of online accessibility, peer-to-peer exchange models, reliance on market liquidity, provision of risk management tools, and engagement in speculative activities related to future events \citep{weidner2022gambling}. These shared features have contributed to the growing popularity of both types of platforms in the digital age \citep{bebbington2017studies}.

\section{The Product}\label{sec3}

\subsection{Financial Products}

In financial markets, there is a wide range of products available to investors, encompassing bonds, stocks, options, and futures. Firstly, bonds are the fixed-income securities, representing debt instruments issued by governments, municipalities, or corporations to raise capital \citep{valdez2016introduction}. When an investor buys a bond, they are essentially lending money to the issuer in exchange for regular interest payments (coupon payments) and the return of the bond's face value at maturity \citep{finnerty1988financial}. For example, Treasury Bills (T-Bills) are short-term U.S. government bonds with a maturity of one year or less, backed by the Treasury Department. T-Bills are considered one of the safest investments in the financial market due to being backed by the full faith and credit of the government \citep{valdez2016introduction}. 

Secondly, stocks (also known as shares or equities) represent ownership in a company \citep{valdez2016introduction}. Stock prices could be volatile and subject to fluctuations based on various factors, such as economic conditions, company performance, industry trends, and investor sentiment \citep{baker2007investor}. 

Thirdly, options and futures are financial derivatives that allow investors to speculate on the price movements of underlying assets without directly owning them. Options are financial contracts between a buyer and a seller, granting the holder the right (but not obligation) to buy or sell an asset at a set price within a defined time frame. The underlying asset can be stocks, commodities, indices and currencies \citep{valdez2016introduction}. There are two main types of options, including call options and put options. Call options give the holders the right to buy the underlying asset, while put options give them the right to sell it, at the strike price before or on the expiration date \citep{hull1993options}. Options can also be categorized by their duration, with short-term options expiring within a year and long-term options (LEAPs) having expirations greater than a year. LEAPs are identical to regular options except that they have longer durations \citep{bakshi2000pricing}. Two other well-known types of options are American options and European options. American options can be exercised at any time between the date of purchase and the expiration date, while European options can only be exercised on their expiration date \citep{hull1993options}. The price of an option, known as the premium, is influenced by factors such as the current price of the underlying asset, the strike price, the time remaining until expiration, market volatility, and interest rates \citep{stutzer1996simple}. The options have various applications, including speculation, hedging, and income generation, making them valuable tools for traders and investors \citep{cuthbertson2001financial}. Compared to options, futures contracts create an obligation for both parties to fulfill the contract on the specified future date \citep{hull1993options}. Futures contracts are commonly used for hedging against price fluctuations in the underlying asset and for speculation, enabling traders to profit from price movements without owning the asset \citep{valdez2016introduction}.  The derivatives provide investors with various strategies to hedge against risk or speculate on price movements, expanding the array of choices beyond traditional stock investments \citep{cuthbertson2001financial}.

\subsection{Gambling Products}

Throughout human history, various forms of gambling and games of chance have been prevalent, including lotteries and sports betting \citep{borna1987gambling}. Lotteries are games of chance where participants purchase tickets or numbers for a chance to win prizes \citep{ariyabuddhiphongs2011lottery}. The odds are typically calculated based on the total number of tickets sold and the number of winning tickets \citep{clotfelter1990economics}. On the other hand, sports betting involves wagering on the outcome of sporting events, with the potential for some degree of skill and analysis to influence betting decisions \citep{hausch1995efficiency}. Compared to lotteries, sports betting offers a wide range of betting choices \citep{labrie2011identifying}. Participants can place bets on various aspects of a sporting event, including the winner, final score, point spreads, and player performance. Sports betting allows for more diverse and strategic betting choices compared to lotteries \citep{hausch1995efficiency}. In the UK, horse racing betting is a popular form of sports wagering, involving predicting race outcomes \citep{huggins2014flat}. However, in peer-to-peer markets including the Betfair betting exchange, participants have the unique opportunity to set their own odds and bet against each other, deviating from the traditional sports betting setup where bookmakers set the odds \citep{casadesus2019platform}.

\subsection{Comparison between Financial and Gambling Products}
There are certainly similarities between products in the financial and gambling markets. While bonds and lotteries have distinct characteristics, they still share some commonalities. First of all, both bonds and lotteries serve as fundraising mechanisms, enabling governments, corporations, or other entities to raise capital from the public \citep{serpeninova2020mapping}. Bonds are a way for these entities to borrow money from investors in exchange for regular interest payments and the return of the principal amount at maturity \citep{valdez2016introduction}, while lotteries involve participants purchasing tickets for a chance to win prizes \citep{ariyabuddhiphongs2011lottery}, thus contributing to the overall funds raised. 

Secondly, both bonds and lotteries entail risks for participants. In the case of bonds, there exists the risk of issuer default on interest payments or failure to repay the principal amount upon maturity \citep{lee2002pricing}. On the other hand, lotteries pose the risk of participants losing the amount spent on tickets, as winning is entirely based on chance, and not all participants can be winners \citep{currie2006risk}. 

For options and sports betting, these two products also share certain similarities, particularly in terms of speculation \citep{borna1987gambling} and duration \citep{mallios2011forecasting}. In options trading, participants speculate on the future price movements of an underlying asset. They can enter into options contracts to buy or sell the asset at a predetermined price within a specified period, based on their anticipation of how the asset's value will change \citep{chang2009informed}. Similarly, sports betting involves participants speculating on the outcome of an event or the performance of a particular team, player, or scenario in a sporting event. Bettors place wagers based on their predictions of the event's outcome, with the potential for winnings or losses depending on the accuracy of their speculation \citep{raney2012chapter}. Both markets involve time-sensitive products, with short-term options and sports bets having brief duration, making short-term price movements more predictable \citep{mallios2011forecasting}.

However, an important difference lies in the economic role of time. In financial markets, maturity is central to valuation, and derivative pricing is inherently time-structured. Models such as Black-Scholes \citep{black1973pricing} and stochastic volatility with jumps \citep{andersen2000jump} can be interpreted as optimal stopping \citep{van1974optimal} or dynamic replication problems, where pricing follows arbitrage-free arguments over continuous time \citep{merton1976option}. By contrast, in traditional gambling markets, bets are typically held until a fixed event date, implying a predetermined resolution and ruling out an optimal stopping interpretation. Peer-to-peer betting exchanges introduce greater flexibility, allowing participants to open and close positions before the end of the event. Although this brings such markets similar to financial trading in operational terms, a fundamental distinction remains. Odds are not derived from replication or no-arbitrage pricing under an equivalent martingale measure; outcomes are externally determined, and there is no maturity-dependent contingent claim structure. Thus, even when time becomes more active, the pricing architecture remains structurally different from that of financial markets.

\section{The Procedure}\label{sec4}

\subsection{Financial Procedures}
The procedures used within the financial market can vary depending on the specific asset being traded and the structure of the market in place \citep{grossman1988liquidity}. The opening price in stock markets is important as it represents the initial trading value of a security at the start of a trading day \citep{valdez2016introduction}. For example, the New York Stock Exchange (NYSE) opens at precisely 9.30 a.m. and the price of the first trade for any listed stock is its daily opening price. This opening price serves as an important marker for that day's trading activity, especially for those interested in short-term outcomes such as day traders \citep{monaghan2009oxidative}. The behaviors of traders in the stock market involve either selling or buying stocks. Selling refers to a recommendation to sell a security or liquidate an asset, while buying entails purchasing a specific security. The bid-ask spread is the difference between the highest price a buyer offers and the lowest price a seller accepts. It reflects market liquidity, with narrower spreads indicating higher liquidity and wider spreads suggesting lower liquidity or higher risk \citep{valdez2016introduction}. Different from stocks, derivatives are financial contracts based on underlying assets, encompassing futures, options, swaps, and forwards \citep{cuthbertson2001financial}. 

\subsection{Traditional Gambling Procedures}
Similar to financial markets, procedures in gambling markets can vary based on the specific products offered. Taking sports betting for example, in traditional gambling markets, participants simply need to choose a reputable sportsbook or bookmaker and place their bets on sports events. Once the bet is placed, they have to wait for the outcome during the events, as the betting odds are fixed in traditional gambling markets \citep{cain2003favourite}. In European gambling markets, two common types of odds are used, including fractional odds and decimal odds \citep{cortis2015expected}. Fractional odds, commonly used in the UK, are represented as a fraction, for instance $2/1$ or $5/2$. The first number in the fraction indicates the potential profit, while the second number represents the stake required \citep{che2017price}. For example, in fractional odds of $2/1$, a participant bets \pounds1, it would potentially win \pounds2 in profit along with its original \pounds1 stake if its bet is successful. On the other hand, decimal odds are expressed as a decimal number, for example $3.00$ or $2.50$. These odds represent the total payout, including both the profit and the original stake \citep{che2017price}. For instance, a participant bet \pounds1 on decimal odds of $3.00$, it would potentially win a total of \pounds3, which includes its original \pounds1 stake and \pounds2 in profit if its bet is successful. An essential aspect of sports betting, especially in horse racing betting, is the starting price. The starting price denotes the initial odds assigned to a horse or team at the beginning of a race or sporting event. This starting price plays an important role in determining potential payouts for winning bets \citep{ruihley2021swiftly}. On-course bookmakers set these prices, but they are influenced by the activities on the betting exchange. Many off-course bookmakers (especially online bookmakers) now offer best odds guaranteed (BOG), which means that if a price is taken at the time of the bet and the starting price is larger, they will pay out the starting price, ensuring that the customer has received the best odds \citep{crafts1985some}. 

Beyond odds quotation formats and payout conventions, the institutional organization of gambling markets differs fundamentally from that of financial markets. A seminal analysis explaining that bookmakers are primarily concerned with balancing their books rather than forecasting true outcome probabilities \citep{levitt2004gambling}. Therefore, odds are adjusted strategically to manage betting imbalances, thereby transferring risk from bookmakers to market participants. As a result, betting odds reflect not only information about event fundamentals, but also institutional objectives and demand-side pressures. Some empirical studies have examined betting market percentages to test the balanced book hypothesis. The literature indicates that wagering volume exerts a systematic influence on odds formation and the dynamic process of market adjustment \citep{levitt2004gambling}. Biased betting distributions can induce systematic distortions in prices, with important implications for market efficiency and behavioral phenomena such as the Hot Hand effect \citep{camerer1989does}. The results suggest that conventional gambling market prices are simultaneously influenced by the information signal and the institutional form of a mechanism. This dual structure separates conventional gambling market structures from financial asset markets, which are primarily determined by competitive valuation and arbitrage dynamics. Consequently, procedural and institutional features must be explicitly considered when applying financial market models to traditional gambling environments.

\subsection{Peer-to-Peer Gambling Procedures}
In contrast to traditional gambling, peer-to-peer gambling markets operate with different procedures, including the back and lay system on the Betfair betting exchange, where individuals can set their own odds \citep{axen2020hedging}. This system allows bettors to either ``back" or ``lay" a selection. When bettors back a selection, they are betting on it to win, while laying a selection means betting against it to win. Betting exchanges are online platforms that enable bettors to bet directly against each other, eliminating the need for intermediaries including bookmakers \citep{vlastakis2009efficient}. For example, Charlie, a sports bettor, enjoys betting on horse racing. After analyzing the recent performances of the runners, Charlie decides to back a particular horse to win the race. They place a bet of 100 pounds on the horse at odds of $2.5$. If this horse wins, Charlie will receive a payout of 250 pounds, which includes its original stake of 100 pounds and 150 pounds in winnings. By backing the horse, Charlie is expressing its confidence in the victory. However, if another horse wins, Charlie will lose its bet. On the other hand, the lay price refers to the price at which a bettor can offer to take bets from others who believe that a specific outcome will not occur \citep{axen2020hedging}. For instance, if one horse is a strong favorite to win a race, the back price might be set at $2/1$ (or $3.0$ in decimal odds). Bettors who believe the horse will lose can offer to ``lay" a bet at a certain price, say $1/1$ (or $2.0$ in decimal odds). This means that if someone takes their bet and the horse loses, the bettors will receive their stake as winnings. However, if the horse does win, they would have to pay out \pounds2, which includes the original stake they wagered plus the additional winnings, to the person who took their bet. The original stake refers to the initial amount of money that a bettor puts at risk when placing a bet \citep{bolen1968gambling}.

Additionally, the back and lay prices on a betting exchange are influenced by the number of bettors willing to back and lay a selection, respectively. As more bettors want to back a horse to win, its odds decrease, while if more bettors want to lay a horse to win, its odds increase. This dynamic and constantly changing market allows users to bet on event outcomes in a way that was not possible with traditional bookmakers \citep{maher1993betting}. This peer-to-peer system creates a more flexible and interactive betting environment for participants.

For peer-to-peer sports betting, there are two different types of betting choices, including pre-game betting and in-game betting. Firstly, pre-game betting is the traditional form of sports wagering and involves predicting the outcome of a game or match before it takes place \citep{gil2007testing}. Secondly, in-game betting, also known as live betting or in-play betting, allows bettors to place wagers during sports events as they unfold in real-time \citep{killick2019play}. Technological advancements have fueled the rise of in-game betting, allowing bettors to dynamically react and place bets based on unfolding events during the event \citep{king2010video}. 

\subsection{Comparison between Financial and Gambling Procedures}
The concept of ``lay” and ``back” in the gambling market share some similarities with call and put options in financial markets \citep{valdez2016introduction}. In the financial markets, both ``bearish” and ``bullish” describe the prevailing sentiment and expectations of market participants and are often used to characterize the overall market direction or the sentiment towards a specific asset \citep{brown2004investor}. The term ``bearish" originates from the way bears attack their prey by swiping downward with their paws, symbolizing the downward movement of prices \citep{lee2002stock}. In financial markets, being bearish implies that investors and traders hold a negative outlook on the market or the asset's future performance. During bearish market conditions, prices are generally expected to decline or remain stagnant. As a result, investors often resort to selling their holdings to avoid potential losses or employ strategies including short-selling to profit from declining prices \citep{lee2002stock}. On the other hand, the term ``bullish" is derived from the way bulls thrust their horns upward, symbolizing the upward movement of prices \citep{lee2002stock}. In financial contexts, being bullish suggests that investors and traders hold a positive outlook on the market or the asset's future performance. In a bullish market, prices are generally expected to rise, prompting investors to buy assets with the expectation of future gains. Both ``lay" and ``put" reflect a bearish sentiment, indicating that the participant believes the value of the asset or the likelihood of an outcome will decrease. On the other hand, ``back” and ``call” show a bullish sentiment, where the participant believes that the value of the asset or the likelihood of an outcome will increase \citep{wr2022sentiment}. 

Additionally, there are similarities between trading in a betting exchange and a stock exchange \citep{arthur2016conceptual}. Firstly, participants can buy and sell assets. In stock markets, these assets are typically company shares, while in peer-to-peer betting markets, they are betting positions on various outcomes of events, including sports matches \citep{noon2013extending}. Secondly, trading in both markets involve matching buyers and sellers to facilitate transactions. The prices of assets are determined by supply and demand, with bids and offers interacting to establish equilibrium \citep{lee2023association}. The primary similarity between bid-ask spread and lay/back odds difference lies in their representation of the difference between two prices associated with a specific item. In the financial markets, the bid-ask spread refers to the difference between the highest price a buyer is willing to pay (the bid price) and the lowest price a seller is willing to accept (the ask price) for a financial asset, including stocks \citep{amihud1986asset}. While in gambling markets, the difference between lay and back odds represents the gap between the odds offered by those who believe an outcome will happen (back odds) and those who believe it will not happen (lay odds) \citep{axen2020hedging}. Both bid-ask spread and lay/back odds difference provide valuable insights into the liquidity and market sentiment related to the respective markets \citep{copeland1983information, wr2022sentiment}. A narrower bid-ask spread or a smaller difference between lay and back odds indicates higher liquidity and market efficiency, where buying and selling or betting on outcomes can occur with minimal price discrepancy \citep{amihud1986asset, wr2022sentiment}. Thirdly, trading in both exchanges can be speculative in nature, with participants attempting to predict the future value of assets to make profitable trades. The speculation is a common strategy utilized by traders seeking to capitalize on potential price movements and generate profits \citep{borna1987gambling}. These similarities highlight the overlap in principles between financial and gambling markets, despite their distinct purposes.

\section{The Participant}\label{sec5}

\subsection{Financial Participants}
Currently, millions of people actively participate in the financial markets, each with their own distinct purpose and strategies to achieve financial gains \citep{allen1997theory}. In traditional financial markets, stock brokers play a significant role as highly skilled professionals. These financial brokers cater to both individual and corporate clients, facilitating investment transactions on their behalf. The stock brokers encompasses various career paths, including stock traders, investment brokers, commodities brokers, and bond brokers \citep{burt2007brokerage}. Additionally, brokers often function as financial advisors, offering guidance to clients on investment portfolios and choices to help them achieve their financial goals \citep{valdez2016introduction}. The investment choices chosen for a client are influenced by their unique financial situation and objectives. For instance, a long-term investor seeking retirement planning advice will likely make different investment decisions compared to an active trader seeking quick returns \citep{barber2001internet}. 

Aside from professional brokers, in financial markets, many speculators willingly take on risks in pursuit of potential profits. Speculators aim to buy assets at low prices and sell them at higher prices. In the case of futures markets, they can also choose to sell first and later buy at a lower price. These speculators actively bet against market movements to capitalize on fluctuations in security prices. Different from speculators, hedgers adopt a different approach in the options market, seeking to minimize risks associated with market uncertainties \citep{valdez2016introduction}. 

Participating in peer-to-peer trading markets is distinct from traditional markets. For instance, to engage in online futures trading on the Shanghai International Energy Exchange in China, investors typically need to open an account with a licensed brokerage firm that provides access to the specific exchange. As a result, participation in peer-to-peer financial markets might be limited to skillful investors \citep{lv2020crude}.

\subsection{Gambling Participants}
In the betting markets, a variety of participants engage in different ways. First of all, casual bettors participate in betting for entertainment. They place bets occasionally and do not invest significant time or effort into studying betting strategies or analyzing odds \citep{platz2001gambling}. Secondly, professional bettors approach betting as a serious endeavor with the aim of making consistent profits from their wagers. These individuals invest a significant amount of time researching and analyzing sports or events, studying statistics, and developing strategies to gain an edge \citep{moore2012self}. Thirdly, another category of participants is high-stakes bettors, characterized by their propensity to place large wagers on events. They are comfortable risking significant amounts of money in the hopes of substantial returns. High-stakes bettors often have larger bankrolls and possess extensive knowledge of the specific sports they are betting on \citep{metz2022high}. Additionally, in peer-to-peer gambling markets, in-play bettors are a distinct group that focuses on placing wagers during an event as it unfolds. They take advantage of real-time odds and adjust their bets based on the evolving circumstances of the game or event. In-play bettors typically have a good understanding of the sport and leverage their ability to assess the momentum and dynamics of the game \citep{hing2018does}. Apart from in-play bettors, spread bettors primarily focus on point spreads in sports betting. Instead of simply predicting the winner, they bet on the margin of victory or defeat. Spread bettors aim to identify discrepancies between the predicted outcome and the bookmakers' line, seeking to exploit favorable spreads \citep{paul2014bettor}. Therefore, the variety of participants in the betting market enriches its complexity. Each group utilizes unique approaches, enhancing the dynamics of betting markets.

\subsection{Comparison between Financial and Gambling Participants}
There are strong similarities between traders in financial markets and bettors in gambling markets. In developed countries, approximately 14\% of trading activities in all financial markets could be categorized as gambling \citep{chen2021searching}. Frequent stock market trading has been interpreted as a manifestation of gambling-like addictive behavior, suggesting that increased ease of trading may encourage speculative participation and heighten behavioral risk for certain investors \citep{mosenhauer2021stock}. In terms of speculators, who actively participate in various forms of gambling, are more likely to engage in buying lottery stocks in the stock markets, indicating conceptual and empirical relationships between speculation and gambling \citep{williams2023financial}. 

The participants in both markets share similar purposes, excluding casual bettors in the gambling markets. Firstly, both markets involve an element of speculation. In financial markets, participants speculate on future asset prices and economic conditions, while in gambling markets, players speculate on the outcomes of games and events \citep{borna1987gambling}. Secondly, participants in both markets rely on information and analysis to make informed decisions. Although both markets may exhibit weak form or semi-strong form informational efficiency \citep{hausch1995efficiency, williams1999information}, financial market participants can get information by news, company reports, and economic indicators \citep{cohen2012corporate}, while gamblers assess odds via historic news, betting analysis report and the matches through media \citep{sagristano2002time}. Thirdly, professional participants in both markets utilize some methods to manage risk and gain profits, for instance, portfolio management in the financial market \citep{grinold2000active} and the draw trading strategy in sports betting \citep{zaloom2004productive}. Finally, some financial participants focus on short-term gains through trading \citep{froot1992herd}, and similarly, some gamblers engage in short-term bets \citep{petry2003patterns}. Both groups maintain a relatively short time horizon, for instance, investors may hold positions for a few minutes or days, aiming to profit from short-term price movements in the financial markets \citep{fischer2018deep}. Similarly, in the Betfair betting exchange, participants place bets on events that typically have immediate or near-term outcomes, including horse racing. During this short-term period, participants can sell and buy their positions or bets according to their professional judgment \citep{slovic2020construction}. These shared characteristics demonstrate the similarities between financial and gambling participants who might use the similar strategies in both fields.

\section{The Strategy}\label{sec6}

\subsection{Financial Strategies}
In order to gain profit in financial markets, participants employ various strategies to mitigate risks, since all investments inherently carry some level of risk \citep{sharpe1999investment}. Firstly, one widely utilized strategy is arbitrage, which capitalizes on price differences for the same asset or security in different markets or locations \citep{hull1993options}. The objective of arbitrage is to earn a profit with risk-free by exploiting these price discrepancies \citep{valdez2016introduction}. The arbitrage is based on the principle of the ``law of one price”, which posits that identical assets should have the same price in an efficient market \citep{protopapadakis1983spot}. However, due to various factors including transaction costs and market inefficiencies, identical assets may not maintain the same price at all times in all financial markets \citep{jones1988transaction}. Taking currency trading for example, if USD/JPY is 146 in Market A and 147 in Market B, the trader buys yen in Market A and sells it in Market B, profiting from the 1-point difference with risk-free. There are various types of arbitrage, including spatial arbitrage, temporal arbitrage and statistical arbitrage. Taking statistical arbitrage as an example, it is a trading strategy that aims to exploit pricing inefficiencies between related financial instruments, including pairs of stocks, based on statistical models and historical price relationships \citep{pole2011statistical}. Different from classical arbitrage, which focuses on identical assets in different markets, statistical arbitrage involves identifying correlated assets or securities that historically move together predictably. Traders using statistical arbitrage develop quantitative models that analyze historical price data and relevant factors to identify potential trading opportunities \citep{pole2011statistical}. These models help them identify instances where the prices of correlated assets deviate from their historical relationship. When the divergence is detected, the trader may take positions to profit from the expected convergence of prices back to their historical relationship \citep{alexander2005indexing}. However, statistical arbitrage also carries some degree of risk and uncertainty. The profitability of these strategies relies heavily on the accuracy of the underlying statistical models and their ability to predict price relationships \citep{krauss2017statistical}. The model-driven statistical arbitrage can assist investors in achieving higher returns \citep{avellaneda2010statistical}. For example, a trader identifies two historically correlated stocks, including Coca-Cola and Pepsi. When their prices deviate, the trader buys the underperforming stock and sells the outperforming one, profiting when the prices converge back to their historical relationship, independent of overall market trends.

Secondly, portfolio management is another important strategy in financial markets. Portfolio management involves the selection, prioritization, and control of an organization's programs and projects in line with its strategic objectives and capacity to deliver \citep{grinold2000active}. The goal is to balance the implementation of change initiatives and the maintenance of business-as-usual, while optimizing return on investment \citep{mathiesen2011100}. The proper project selection could enhance the efficiency of portfolio management \citep{valdez2016introduction}. 

Alongside arbitrage and portfolio management, other popular strategy in financial markets includes risk management \citep{valdez2016introduction}. Each of these strategies serves an unique purpose in helping participants navigate the complexities of financial markets and work towards their investment objectives.

\subsection{Gambling Strategies}
Since there is a strong relationship between gambling and speculation in the financial markets \citep{borna1987gambling}, the strategies employed in traditional gambling markets are different from those in the financial markets. In traditional sports betting, for example horse racing, the odds are fixed and do not change during the course of the match \citep{labrie2007assessing}. Bettors are unable to buy and sell their bets during the match, thus they may utilize the cross-track betting strategy, which allows them to place wagers on horse races taking place at different racetracks or venues \citep{hausch1990arbitrage}. This strategy enables bettors to access race information and odds from multiple tracks and place their bets accordingly, even if they are not physically present at those locations. 

However, different from traditional gambling markets, in peer-to-peer betting exchanges, bettors have the flexibility to buy and sell the back, lay, or draw bets during the sports event \citep{casadesus2019platform}. Therefore, statistical arbitrage can be applied during trading, which involves exploiting discrepancies in the odds set by different bookmakers or exchanges \citep{vlastakis2009efficient}. By strategically placing bets on all possible outcomes of an event in a way that guarantees a profit regardless of the actual outcome, bettors can make money without taking on any risk. This strategy is often achieved by identifying situations where the odds offered by different bookmakers or exchanges are inconsistent, allowing bettors to place bets that are certain to result in a profit \citep{forrest2012threat}. For example in a basketball betting, with odds of 3.0 for Team X win at Bookmaker A and 3.2 for Team Y win at Bookmaker B, the bettor ensures a guaranteed profit by covering all outcomes. Another valuable strategy that could be used in peer-to-peer gambling markets is the practice of backing at high and laying at low prices. This strategy involves capitalizing on price movements in the betting markets. When bettors back a selection, they aim to place their bet at higher odds than those available later. Conversely, when they lay a selection, they seek to do it at lower odds than what may be available later \citep{oner2012inflation}. This strategy is to generate profits by making opposing bets at different times. When the odds for a selection shorten, bettors can lay it at a lower price than what they initially backed it for, guaranteeing a profit regardless of the outcome \citep{nordsted2010mastering}. On the other hand, if the odds lengthen, they can back the selection at a higher price than what they initially laid it for, once again securing a profit irrespective of the outcome \citep{graham2005intelligent}. Implementing this strategy effectively requires vigilantly monitoring odds and promptly executing trades to seize favorable price fluctuations. Apart from this strategy, the ``lay the draw" trading strategy \citep{zaloom2004productive} is commonly employed in sports betting, particularly in soccer. It involves placing a lay bet on the draw outcome of a soccer match before the game starts, and then backing the draw at longer odds during the match. The goal is to lock in a profit by taking advantage of shifts in the odds during the game. Thus, the strategies used in traditional gambling markets and peer-to-peer betting exchanges differ due to the fixed odds nature of traditional sports betting and the dynamic betting opportunities in peer-to-peer exchanges.

\subsection{Market Efficiency in Gambling Markets}
Market efficiency has long served as a central benchmark in financial economics and has increasingly been applied to betting markets, where odds are commonly interpreted as collective beliefs about uncertain outcomes. \citet{zuber1985beating} provided one of the earliest systematic examinations of efficiency in horse racing markets and reported evidence suggesting informational inefficiency in betting prices. \citet{sauer1988hold}, however, re-evaluated these findings and argued that betting markets were largely consistent with weak-form efficiency once sampling procedures and methodological issues were appropriately addressed. This academic debate played a pivotal role in legitimizing betting markets as an empirical setting for testing the Efficient Markets Hypothesis.

Some empirical studies have reported certain pricing regularities, such as the favourite-longshot bias \citep{oikonomidis2013weak}. Nevertheless, the economic significance of these effects remains limited, and their exploitability is often substantially reduced after accounting for transaction costs and market microstructure constraints. Consequently, betting markets are commonly regarded as approximately efficient in practice \citep{sauer1988hold}.

The emergence of peer-to-peer betting exchanges has further strengthened the structural parallels between betting and financial markets. In these environments, odds are no longer set solely by bookmakers but are instead formed through continuous trading interactions among participants \citep{sauer1998economics}. This price discovery mechanism closely resembles order-driven financial markets and reinforces the relevance of efficiency-based analysis. Consequently, considerations of market efficiency provide a natural theoretical foundation for the application of financial trading strategies and portfolio allocation frameworks in gambling contexts. Within this setting, quantitative betting strategies can be interpreted as attempts to exploit transient deviations from an otherwise approximately efficient market.

\subsection{Pricing Structure and Price Formation Mechanisms}
Different from financial markets, where pricing equations are typically derived from no-arbitrage conditions and intertemporal optimization, gambling markets do not admit an equivalent martingale framework or replication-based valuation. In finance, asset pricing, particularly for derivatives, is built on stochastic discount factors, risk-neutral measures, and dynamic hedging, with time and state-contingency explicitly embedded in the pricing structure \citep{andersen2000jump}. By contrast, gambling prices are outcome-based rather than replication-based. Odds are shaped by supply–demand imbalances, bookmaker margin considerations, or peer-to-peer order flow, rather than by arbitrage-free valuation relative to traded assets \citep{labrie2007assessing}. As a result, pricing in gambling markets cannot generally be expressed as a no-arbitrage partial differential equation or as a risk-neutral expectation of discounted payoffs. Instead, it reflects market-clearing probabilities, participant behavior, and microstructure effects. This distinction implies that, even when time is incorporated explicitly, the mathematical structure of pricing in gambling markets differs systematically from that in financial markets \citep{newall2021structural}.

Gambling markets are organized differently from financial markets largely because their price adjustment mechanisms differ substantially \citep{levitt2004gambling}. In traditional bookmaker markets, odds are adjusted infrequently and strategically. Price revisions are not driven solely by information arrival but also by the bookmaker’s objective of managing exposure and balancing the book. As a result, price dynamics are partially administratively determined rather than fully market-clearing. However, this characterization does not fully apply to peer-to-peer betting exchanges. In peer-to-peer gambling platforms, such as order-driven betting markets, odds are determined through continuous interaction between back and lay orders submitted by participants. Therefore, price adjustment occurs endogenously through order flow, resembling the decentralized price discovery process observed in financial exchanges. As a consequence, peer-to-peer gambling markets show greater market-based price formation levels than bookmaker systems.

Nevertheless, even in peer-to-peer environments, a fundamental structural difference remains. Financial asset prices are typically derived within no-arbitrage frameworks that admit intertemporal replication and risk-neutral valuation. By contrast, betting odds, whether set by bookmakers or determined through exchange order flow, represent equilibrium-implied probabilities of mutually exclusive outcomes rather than prices of dynamically hedgeable claims. Therefore, in spite of similar trading mechanics, the underlying structural pricing architecture is systematically different from that of financial markets \citep{hegarty2026market}.

\subsection{Comparison between Financial and Gambling Strategies}
Building upon the efficiency properties shared by financial and betting markets, a meaningful comparison of trading and betting strategies can be established. First of all, significant arbitrage opportunities exist in the cross-track betting strategy \citep{hausch1990arbitrage}. Similarly, in the financial markets, arbitrage involves exploiting price discrepancies between different assets or securities to secure risk-free profits \citep{valdez2016introduction}. The goal of arbitrage in both cases is to take advantage of price differences and earn a profit without taking on any risk. The statistical arbitrage strategy is found to be effective in the financial market and certain practices in the gambling market, including Betfair betting exchange \citep{vlastakis2009efficient}. In the financial markets, statistical arbitrage involves using quantitative models and algorithms to identify trading opportunities based on statistical patterns and anomalies in asset prices \citep{alexander2005indexing}. Similarly, in the Betfair betting exchange, bettors can use statistical analysis to identify discrepancies in odds offered by different participants and place bets that ensure a guaranteed profit regardless of the outcome. 

Secondly, diversification plays an important role in both cross-track betting and portfolio management. In cross-track betting, bettors diversify their bets by placing wagers on multiple races at different tracks \citep{hausch1990arbitrage}. This spreading of bets helps bettors manage risk and potentially increase the chances of gaining rewards. Similarly, in portfolio management, investors diversify their investments across various asset classes or securities to reduce the impact of individual asset risks and optimize returns \citep{bekaert1998distributional}. By diversifying their portfolios, investors aim to achieve a balanced risk-return profile. Recent research has further reinforced this parallel by framing sports betting within a financial asset perspective. Betting positions are increasingly viewed as portfolio components whose returns and risks can be formally evaluated using financial risk management principles. For instance, sports betting portfolios exhibit risk characteristics comparable to conventional financial assets, thereby supporting the application of diversification-based portfolio allocation frameworks in betting environments \citep{moskowitz2021asset}. 

In addition to arbitrage-based and optimal staking strategies, a complementary strand of the literature has examined behavioral and momentum-driven patterns in betting markets. A classic example is Hot Hand hypothesizing that market players tend to attribute recent success toward future performance and lead to systematically trend-following behavior \citep{camerer1989does}. \citet{camerer1989does} provided early empirical evidence that such beliefs influence betting decisions in sports markets, demonstrating that bettors systematically overreact to recent winning streaks. \citet{brown1993does} further investigated this phenomenon and showed that betting volumes and odds adjustments often reflect momentum-driven perceptions rather than purely fundamental information. Hot Hand effects are persistent but context-dependent, varying across sports, market structures, and levels of market liquidity \citep{camerer1989does}. These findings closely parallel momentum trading in financial markets, where past returns influence investor expectations and trading behavior \citep{helfat2009dynamic}. The Hot Hand literature highlights that betting markets are shaped not only by informational efficiency and arbitrage forces, but also by behavioral biases and adaptive learning mechanisms. This behavioral perspective therefore complements efficiency-based and arbitrage-oriented frameworks, providing a more comprehensive understanding of trading dynamics in gambling markets. 

Therefore, the strategies used in both financial and gambling markets share common elements, including arbitrage opportunities, diversification for risk management, and the use of statistical analysis to identify profitable opportunities. While the contexts may be different, the underlying principles of strategic decision-making remain similar in these two markets. As trading-oriented financial and gambling markets are closely related, the transferability of financial theories is highly contingent on institutional structure. Peer-to-peer betting platforms in particular mirror order-driven financial exchanges where prices emerge endogenously from participant interaction. In these environments, we can make a meaningful translation of our ideas including statistical arbitrage, portfolio allocation, and risk-adjusted performance measurement. But traditional bookmaker markets are governed by administratively set odds; prices are adjusted strategically in order to cope with exposure rather than by continuously clearing a market. Collectively, the reviewed studies suggest that while institutional mechanisms differ, the underlying trading logic exhibits structural similarities. This realization strongly underpins our argument that peer-to-peer gambling markets can be thought of essentially as quasi-financial trading environments, but subject to relevant boundary conditions.

\section{Conclusion}\label{sec7}
As demonstrated above, there are striking resemblances between financial and gambling markets. The reviewed literature provides consistent evidence that although financial and gambling markets differ in institutional design and price adjustment mechanisms, peer-to-peer gambling platforms share structural features with order-driven financial markets. The convergence identified across the five analytical dimensions substantiates our overarching thesis that cross-market strategy transfer is feasible under clearly defined institutional conditions. Firstly, the functions of financial exchanges bear a resemblance to those of online betting exchanges, for example Betfair betting exchange \citep{casadesus2019platform}. 
Secondly, certain financial products retain the element of speculation akin to sports betting, as seen in options trading \citep{chang2009informed, raney2012chapter}. 
Thirdly, in both peer-to-peer financial and betting markets, participants can buy and sell the products based on the bid-ask spread or lay/back odds difference \citep{copeland1983information}.  
Finally, certain strategies, such as statistical arbitrage \citep{krauss2017statistical, vlastakis2009efficient}, are effective in generating gains in both markets.

Given these similarities, it is reasonable to explore the potential application of successful financial models and strategies in the gambling markets. By leveraging the knowledge and expertise from the financial field, gamblers and investors in the gambling markets may be able to enhance their decision-making processes and achieve more favourable outcomes. However, it is essential to recognize that gambling inherently carries risks, and any strategies applied should be done with careful consideration and understanding of the specific nuances and regulations governing the gambling industry. While financial models can help to drive quantitative approaches on gambling exchanges, this cannot translate directly into structural pricing equations based on no-arbitrage replication. However, gambling prices emerge from probability aggregation and exposure management rather than from intertemporal equilibrium pricing. The analysis suggests that the applicability of financial market models to gambling markets is conditional rather than universal. Peer-to-peer betting platforms exhibit structural features that permit the adaptation of trading and portfolio-based strategies. However, structural pricing equations derived from replication, optimal stopping, or continuous-time no-arbitrage frameworks remain intrinsically tied to the institutional architecture of financial markets. Recognizing these boundaries clarifies both the potential and the limitations of cross-market model transfer. 

Although these similarities indicate substantial structural convergence, important differences persist in information asymmetry, regulatory regimes, and behavioral incentives. Such distinctions constrain the direct transferability of financial models to gambling markets and underscore the need for market-specific adaptation. Accordingly, a unifying analytical framework is required to delineate the conditions under which financial concepts and strategies can be meaningfully applied to gambling contexts. Within this framework, both financial trading and gambling may be formalised as portfolio selection problems over distinct outcome spaces. Betting odds may be interpreted as asset prices, wager allocations as portfolio weights, and realized outcomes as asset payoffs. This representation enables established financial tools, such as arbitrage analysis, risk-adjusted performance evaluation, and growth-optimal strategies, to be systematically reformulated for application in betting markets. This formulation establishes gambling markets as quasi-financial trading environments, providing a coherent conceptual basis for the application of formal risk management and portfolio allocation principles. The findings further suggest practical and policy implications for market transparency, participant protection, and platform design. Future research may extend this framework to multi-asset betting portfolios, behavioral dynamics, and empirical tests of financial model transferability. Therefore, this study contributes to a more integrated understanding of speculative decision-making at the intersection of finance and gambling.

\section*{Declarations}
\subsection*{Ethical Approval}
This article does not contain any studies with human participants performed by any of the authors.
\subsection*{Conflict of interest}
Not applicable.
\subsection*{Availability of Data and Material}
The authors confirm that the data supporting the findings of this study are available within the article and its supplementary materials.
\subsection*{Author's contribution}
Dr. Valentin Popov and Haoyu Liu developed the theoretical formalism. Haoyu Liu contributed to the final version of the manuscript. Dr. Carl Donovan and Dr. Valentin Popov supervised the project.
\subsection*{Acknowledgment}
Haoyu Liu gratefully acknowledges funding from the China Scholarship Council and editorial support from Prof. Len Thomas, Dr. Benjamin Baer and Dr. Paula Villegas Verdu.

\nolinenumbers
\newpage
\begin{spacing}{0.1}

\bibliography{sn-bibliography}
\end{spacing}

\end{document}